\documentclass[pre,aps,twocolumn,floatfix,superscriptaddress,amsmath,amssymb,showkeys,nofootinbib,10pt]{revtex4-2}
\usepackage[english]{babel}
\usepackage[utf8]{inputenc} 
\usepackage{xcolor}
\usepackage{graphicx}
\usepackage{tabularx}
\usepackage{hyperref}
\hypersetup{colorlinks=true, linktoc=all, linkcolor=blue, linktocpage, citecolor=blue}

\newcommand{\be}{\begin{equation}}\newcommand{\ee}{\end{equation}}
\newcommand{\bea}{\begin{eqnarray}}\newcommand{\eea}{\end{eqnarray}}
\newcommand{\brr}{\begin{array}}\newcommand{\err}{\end{array}}
\newcommand{\bit}{\begin{itemize}}\newcommand{\eit}{\end{itemize}}
\newcommand{\ben}{\begin{enumerate}}\newcommand{\een}{\end{enumerate}}

\newcommand{\bbm}{\begin{bmatrix}}\newcommand{\ebm}{\end{bmatrix}}
\newcommand{\ba}{\begin{array}}
\newcommand{\ea}{\end{array}}

\newtheorem{mydef}{Definition}
\newtheorem{Lemma}{Lemma}
\newcommand{\bd}{\begin{mydef}} \newcommand{\ed}{\end{mydef}}
\newcommand{\bthe}{\begin{theorem}} \newcommand{\ethe}{\end{theorem}}
\newcommand{\ble}{\begin{Lemma}} \newcommand{\ele}{\end{Lemma}}

\def\ha{\frac{1}{2}}

\def\lf{\left}

\def\ri{\right}
\def\al{\alpha}

\def\1{{_{1}}}\def\2{{_{2}}}

\def\noHe0{:\;\!\!\;\!\!:H_e(0):\;\!\!\;\!\!:}
\def\noHm0{:\;\!\!\;\!\!:H_\mu(0):\;\!\!\;\!\!:}

\def\lf{\left}

\def\ri{\right}

\def\al{\alpha}

\def\1{{_{1}}}\def\2{{_{2}}}

\begin{document}
\title{Coarsening in the long-range Persistent Voter Model}

\author{Jeferson J. Arenzon}
\email{arenzon@if.ufrgs.br}
\affiliation{Instituto de Física, Universidade Federal do Rio Grande do Sul, 91509-900, Porto Alegre - RS, Brazil} 
\affiliation{Instituto Nacional de Ciência e Tecnologia - Sistemas Complexos, Rio de Janeiro RJ, Brazil}

\author{F. Corberi}
\email{fcorberi@unisa.it}
\affiliation{Dipartimento di Fisica, Universit\`a di Salerno, Via Giovanni Paolo II 132, 84084 Fisciano (SA), Italy}
\affiliation{INFN Sezione di Napoli, Gruppo collegato di Salerno, Italy}

\author{W. G. Dantas}
\email{wgdantas@id.uff.br}
\affiliation{Departamento de Ciências Exatas, EEIMVR, Universidade Federal Fluminense, CEP 27255-125, Volta Redonda - RJ, Brazil}

\author{L. Smaldone}
\email{lsmaldone@unisa.it}
\affiliation{Dipartimento di Fisica, Universit\`a di Salerno, Via Giovanni Paolo II 132, 84084 Fisciano (SA), Italy}
\affiliation{INFN Sezione di Napoli, Gruppo collegato di Salerno, Italy}

\begin{abstract}
We investigate the coarsening kinetics in a long-range variant of the Persistent Voter Model in space dimensions $d=1$ and 2. In this model, agents can hold two confidence levels, normal and zealot. Normal agents imitate another opinion chosen at a distance $r$ with probability $P(r) \propto r^{-\alpha}$, with $\alpha >d$. On the contrary, while in the zealot state, agents keep their own opinion. Normal (zealot) agents can become zealots (normal) if their opinion is equal (different) to that of the chosen neighbour. Through numerical simulations we show that, for any values of $\alpha$, the model belongs to the same universality class of the long-range Ising model quenched to a small (non-zero) temperature, similarly to what was already known for the nearest-neighbor case. For the one-dimensional case, we further develop an analytical treatment, which reproduces the $\alpha$-dependence of the correlation length and the functional form of the correlation function. These results not only confirm that the introduction of opinion inertia mitigates the strong interfacial noise present in the Voter model, thus reinstating the basic kinetic mechanism of the Ising model, but also expand the applicability of this correspondence.
\end{abstract}

%\keywords{first keyword, second keyword, third keyword}
\maketitle

%%%%%%%%%%%%%%%%%%%%%%%%%%%%%%%%%%%%%%%%%%%%%%%%%%%%%%%%%%%%%
\section{Introduction}
Coarsening is a defining feature of systems undergoing an ordering process~\cite{Bray94}, and arises in contexts as diverse as liquid crystals~\cite{SiArDiBrCuMaAlPi08,Almeida23,AlAr25}, superconductors~\cite{PrFiHoCa08}, and a variety of biological phenomena~\cite{Mcnally17,Smerlak18,Siteur22,Grober23}, among many other examples~\cite{Scheucher88,Cox86}.
In such systems, the space is partitioned into domains whose typical size grows over time.
Within the Model A universality class, scalar, non-conserved order parameter dynamics, the scaling regime is characterized by a single growing length scale, such as the average domain radius, which usually increases as ${\cal L}(t) \sim t^{1/2}$ in systems with short range interactions.

%This scaling behavior is captured both by systems with an interfacial energy cost, such as the quenched Ising model (IM), and by systems without such a cost, e.g., the Voter model (VM)~\cite{HoLi75,Redner19}.
This scaling behavior is captured by systems both with or without an interfacial energy cost such as, e.g., the quenched Ising model (IM)~\cite{Bray94,ArBrCuSi07} and the Voter model (VM)~\cite{HoLi75,Redner19,CaFoLo09}, respectively.
These two models represent important limiting cases of the system studied here.
Albeit equivalent in one dimension, for $d>1$ their domain morphologies are qualitatively different due to the distinct mechanisms driving interface evolution: curvature-driven surface tension in the IM versus interfacial noise in the VM.
As a result, in $d=2$ the VM sustains rough interfaces, and the fraction of active sites (those at the interfaces between opposite states) has a much slower decay than in Ising-like systems~\cite{DoChChHi01}.
For $d \geq 3$, surface noise halts coarsening altogether for the VM~\cite{corberi2024coarsening} while for the IM~\cite{Lipowski99,SpKrRe01a,SpKrRe01b,OlKrRe11a,ArCuPi15}, the growth is interrupted at very low temperatures due to the large number of flat interfaces.

The original VM~\cite{HoLi75}, where spins on a lattice adopt the state of a randomly chosen nearest neighbor (nn), offers a minimal description of opinion dynamics approaching an absorbing consensus state in which one opinion eventually dominates.
Numerous generalizations have been explored, including intermediate states between the two original opinions~\cite{VoRe12,VeVa18,VaKrRe03,LaSaBl09,WaLiWaZhWa14,BaPiSe15,SvSw15}.
Such intermediate states may represent complete indecision or more nuanced, moderate viewpoints that, together with other mechanisms that slow down state changes~\cite{StTeSc08,StTeSc08b,PeKhRo20,ViLlToAn24}, introduce inertia in the switching process.
A common outcome of several of these variants is the emergence of an effective surface tension, which drives the system toward consensus with power-law scaling that is similar, though not identical, to the IM.
The original VM is exactly solvable in any dimension, lattice, or interaction form~\cite{FrKr96,BeFrKr96,Corberi_20241d,corsmal2023ordering,corberi2024coarsening,corberi2024aging,Corberi_2024}, owing to the linearity of the moment equations for its probability distribution.
In contrast, because variants with intermediate states are usually analytically harder (except in special cases that can be mapped onto general field-theoretic descriptions~\cite{DaGa08}), most of the insights come from numerical simulations.  

In the Persistent Voter Model (PVM), introduced in Refs.~\cite{LaDaAr22,LaDaAr24}, each opinion can be held with two confidence levels: normal or extreme (zealots).
Zealots are agents that resist opinion changes regardless of neighbors influence~\cite{Mobilia03,MoPeRe07,GaJa07,MaGa13,CoCa16}, a persistence that here is transient because confidence can be reset, returning agents to the normal voter state.
Such transient zealotry, known to optimize the time to consensus in some cases, mirrors critical thinking~\cite{AmAr18,AmDaAr20}, where a minimum amount of evidence is necessary to turn an opinion.
In opposition, while a normal voter is too volatile, changing its opinion with much less information available, the permanent zealot is an irrational agent whose opinion does not change despite the evidences.
Such stubbornness may also be implemented through agents that revert to their original opinion at random times~\cite{Grange23}.
Since the transition from a normal voter to a zealot depends on a continuated built up of confidence, i.e., on the past states of an agent, the general PVM formulation requires memory skills from the agents.
However, a simplified variant has been recently introduced~\cite{ourpaper}, demonstrating that the most distinctive features  of the PVM do not require such explicit memory.
Concentrating on 1D and 2D lattices with short range interactions, we have shown that the density $\rho(t)$ of active sites decays with the same power-law exponent as in the Model A universality class to which the low-temperature IM belongs, i.e., $\rho(t)\sim t^{-1/2}$.

\begin{figure}[htb]
\includegraphics[width=0.7\columnwidth]{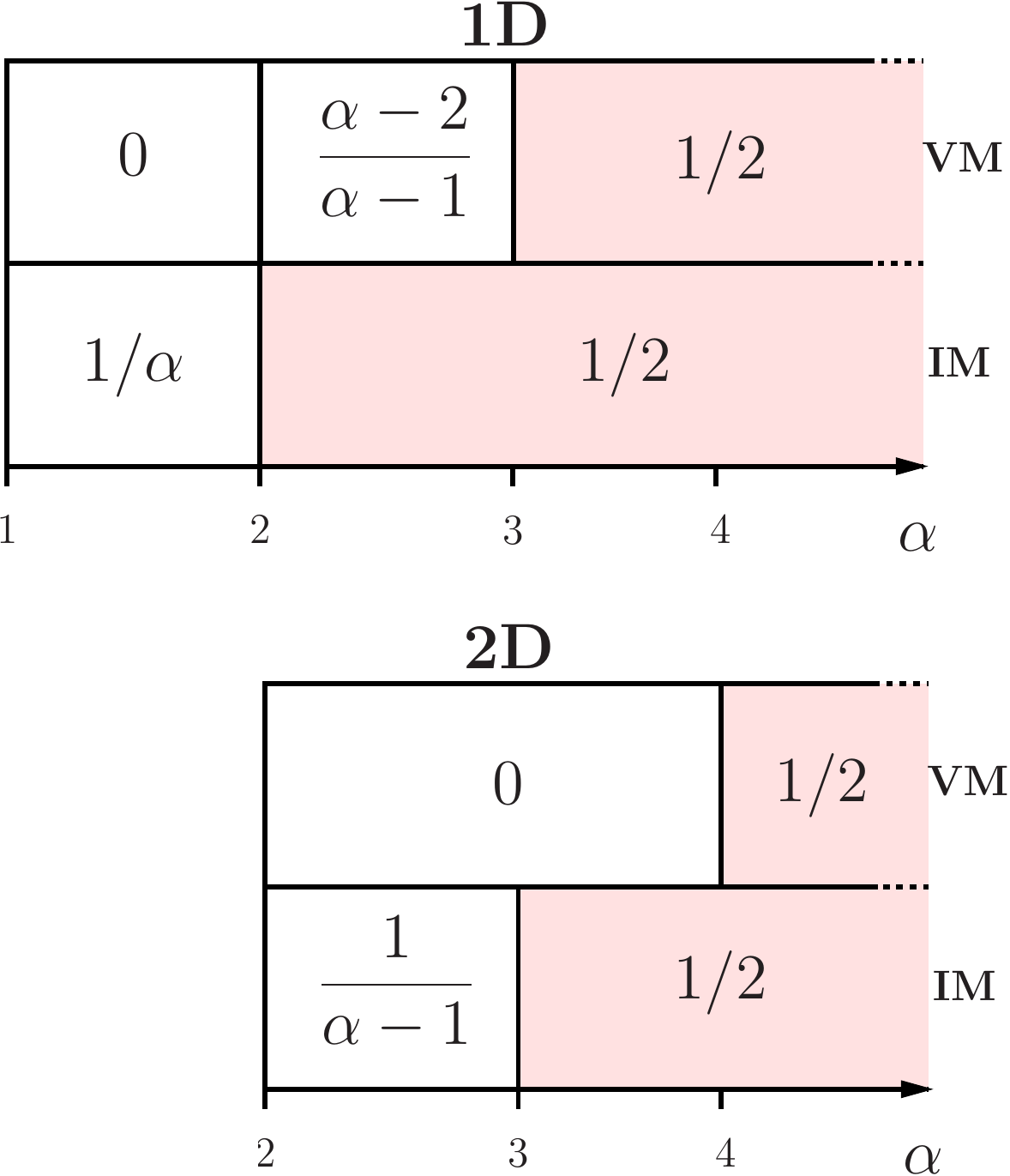}

\caption{Long-range universality classes, given by the asymptotic value of the exponent $1/z$ in Eq.~(\ref{eq.rhot}), as a function of the long-range exponent $\alpha>d$, for the VM and the IM quenched to a low (but finite) temperature~\cite{PhysRevE.49.R27,PhysRevE.50.1900,PhysRevE.99.011301} for both $d=1$ (top) and $d=2$ (bottom). Notice that in all cases there is a value of $\alpha$, $\alpha_{\mathrm{SR}}$, above which we recover the short-range exponent 1/2 (red regions in the figure).} 
%In the 1D IM, the full $\alpha$-dependence is $z=2$ for $\alpha>2$ and $z=\alpha$ for $1< \alpha \le 2$.}
\label{1suzd1and2}
%\vspace{1cm}
\end{figure}

In recent years, interest in long-range nonequilibrium statistical models has grown considerably, with attention focused on phase-ordering kinetics~\cite{PhysRevE.49.R27,PhysRevE.50.1900,PhysRevE.99.011301,Corberi_2017,Corberi2019JSM,PhysRevE.103.012108,Corberi2021SCI,CORBERI2023113681,Corberi2023PRE,PhysRevE.102.020102,Corberi_2026} and other aspects~\cite{campa2009statistical,book_long_range,DauRufAriWilk}.
In particular, the coarsening kinetics of the  VM~\cite{Corberi_20241d,corsmal2023ordering,corberi2024coarsening,corberi2024aging,Corberi_2024} and the IM~\cite{corberi2019one,PhysRevE.49.R27,PhysRevE.50.1900}, both with pairwise interactions decaying as $r^{-\alpha}$, $\alpha>d$, have been widely investigated. 
%When such systems are quenched to low temperature, 
For the long-range version of both models, the interface density decays as
\begin{equation}
\rho(t) \propto t^{-1/z} , 
\label{eq.rhot}
\end{equation}
where the exponent $1/z$ generally depends on $\alpha$ (see Fig.~\ref{1suzd1and2}) and falls within these two regimes:
\begin{itemize}
\item
Short-range (SR) regime: For $\alpha > \alpha_{\mathrm{SR}}$, the dynamics is similar to that of nn models ($z=2$). 
\item
Weak long-range (WLR) regime: For $d < \alpha \le \alpha_{\mathrm{SR}}$, $z$ generally depends on $\alpha$. In the VM there is also an interval with $1/z=0$, where the system gets trapped in disordered {\it metastable} states, which occurs for $\alpha \le \alpha ^*(d)$, with $\al^*(d\leq 2)=2d$ %$\al^*(2)=4$ 
and $\al^*(d\ge 3)=\infty$. The same does not occur in the IM universality class. 
\end{itemize}
The crossover values $\alpha=\alpha_{SR}$ and $\alpha=d$ are associated to peculiar properties~\cite{Dyson1969,Thouless1969} of the system and logarithmic corrections ~\cite{Corberi_20241d,corsmal2023ordering,corberi2024coarsening,corberi2024aging,Corberi_2024}. We will not consider these special cases in this paper.

Here we investigate the effect of including interactions with distant agents in the PVM.
%Because the model has an emergent Ising-like behavior, in order to compare both models we concentrate on the regime with $\alpha >d$ since otherwise non-extensivity and  non-additivity may drastically change the properties. 
The presence of zealots in the PVM offers the possibility to model the formation of extremists and, in this work, the differential role of distant and close contacts.
As detailed in the next section, a focal agent and a neighbor at distance $r$ are chosen and the former, if not a zealot, may change its opinion or become a zealot depending on the opinion of its neighbor.
On the other hand, if it is a zealot it either remains in this state or become a normal voter.
Two different scenarios are considered.
First, the unrestricted case, no constraint is imposed on the distance $r$.
In the second case, the transition to and from the zealot state is restricted to be a consequence of the interactions with nearest-neighbors ($r=1$) while opinion changes remain distance-independent.

We perform simulations for several values of $\alpha$ showing that, in both scenarios, the PVM falls within the universality class of the IM with analogous long-range interactions, i.e. with a coupling constant $J(r)$ among spins decaying algebraically with distance as $r^{-\al}$. 
However, the unrestricted case presents strong finite-size effects and a slower convergence to the asymptotic regime when $d<\alpha\leq \alpha_{\mathrm{SR}}$. 
We also provide an analytical treatment in 1D, computing the two-point correlation functions and the correlation length for any value of $\al$.
The paper is organized as follows: in Sec.~\ref{model} we define the model and the transition probabilities for normal voters and zealots; in Sec.~\ref{simul} we present the simulation results; in Sec.~\ref{sec.correlators} we derive analytical results for the 1D case. Finally, in Sec.~\ref{conc} we summarize our conclusions.

%%%%%%%%%%%%%%%%%%%%%%%%%%%%%%%%%%%%%%%%%%%%%%%%%%%%%%%%%%%%
\section{The model} \label{model}

We consider $N=L^d$ agents located on a $d$-dimensional square lattice of linear size $L$.
In the VM with long-range interactions~\cite{Corberi_20241d,corsmal2023ordering,corberi2024coarsening,corberi2024aging,Corberi_2024}, a randomly selected agent (spin) $S_i$ at site $i$ adopts the state $S_j$ of another agent at site $j \neq i$, chosen according to the probability distribution
\begin{equation}
P(r) \ = \ \frac{r^{-\alpha}}{Z} , 
\label{eq.pr}
\end{equation}
where $r$ is the distance between $i$ and $j$, and
\be
Z \ = \ \sum_p n_p\,\ell_p^{-\alpha} , 
\label{eqZ}
\ee
where $p$ is an integer labeling the possible distances $\ell _p$ (e.g., in $d=2$, $\ell_p=1,\sqrt 2 ,2,\sqrt 5,\ldots$ for $p=1,2,3,4,\ldots$), and $n_p$ is the number of sites located at distance $\ell_p$ from a given site. 
In the nn VM, Eq.~(\ref{eq.pr}) reduces to $P(r) = \zeta^{-1}\delta_{r,1}$, where $\zeta$ is the coordination number.

The PVM~\cite{LaDaAr22,LaDaAr24} introduces inertia into the VM: an agent with sufficiently strong confidence in its opinion (a zealot) does not change state.
Whenever two agents with the same opinion interact, their confidences $\eta_i$ and $\eta_j$ increase by a positive increment $\Delta\eta$
and must grow over multiple interactions before reaching the zealot threshold $\eta_i \geq 1$.
In the VM, domain growth is driven by interfacial noise, and domain boundaries are rough.
In contrast, in the PVM, zealots tend to form inside the domains, constraining the normal voters to accumulate near the interfaces in a  boundary layer whose width depends on $\Delta\eta$.
In the particular version of the PVM that we consider here, this region is very thin (a few lattice sites) and the interfaces are smoother, a characteristic of an effective surface tension. Following the definitions in Ref.~\cite{OlMeSa93}, there is still no bulk noise in the model but the zealots reduce the interfacial noise while inducing an emergent curvature-driven domain growth, increasing the similarity with the IM.

In Ref.~\cite{ourpaper}, a simplified nn PVM was proposed that still reproduces this key feature of the original model, i.e. the emergent IM-like dynamics.
Differently from the original PVM, in this simplified version only the confidence of the focal agent is affected by the exchanges between neighbors.
Moreover, for $\Delta\eta = 1$, the  confidence $\eta_i$ may be replaced by a binary variable $\theta_i$ that is $+1$ for zealots and $-1$ for normal voters.
A normal voter becomes a zealot after a single interaction with a like-minded neighbor ($S_i = S_j \ \Rightarrow\ \theta_i = 1$), and reverts to a normal voter after interacting with an opponent ($S_i \neq S_j \ \Rightarrow\ \theta_i = -1$).
In the present work we only deal with this specific version of the PVM.
The probability for a zealot to change its status in the nn PVM is
\be
\label{eq.wti}
w(\theta_i) \ = \ \frac{1}{2}\left(1-\frac{S_i\theta_i}{2d}\sum_{\delta}S_{i+\delta}\right) , 
\ee
where the sum runs over the nn of site $i$. This remains valid in the restricted scenario mentioned above for which analytical results will be obtained (in the unrestricted case, results from numerical simulations will be presented instead).

In the long-range VM, the flipping probability for spin $S_i$  is~\cite{Corberi_20241d,corsmal2023ordering,corberi2024coarsening,corberi2024aging,Corberi_2024}
\be \label{wlrv}
w_{\rm LR}(S_i) \ = \ \frac{1}{2} \sum _p P(\ell_p) \sum _{k=1}^{n_p}(1-S_iS_k) ,
\ee
where $k$ labels the $n_p$ sites at distance $\ell_p$ from $i$.
In both scenarios discussed above, opinions follow a long-range PVM
%while the confidence level can change only through nn interactions. Because of that, we retain Eq.~\eqref{eq.wti}, but modify 
and Eq.~\eqref{wlrv} is modified to account for zealots:
\be
\label{eq.wsi}
w(S_i) \ = \ \frac{1}{2}\left(\frac{1-\theta_i}{2} \right) \sum _p P(\ell_p) \sum _{k=1}^{n_p}(1-S_iS_k) . 
\ee
%As we will show, defined in this way the model behaves exactly as a natural long-ranged extension of the nn Markovian PVM. In particular, 
Reference~\cite{ourpaper} showed that the nn version of the simplified PVM belongs to the IM universality class, rather than the VM class.
This will be expanded here to include its long-range counterpart as well.
Moreover, as discussed in Sec.~\ref{sec.correlators}, the restricted case where $\theta_i$ is updated only for $r=1$ remains analytically tractable. 
%Finally, we have checked, at least in $D=1$, that adding long-range also to the zealots prevents the system from ordering for any $\alpha$.
%In the IM, the full $\alpha$-dependence of the dynamical exponent is $z=2$ for $\alpha>2$ and $z=\alpha$ for $1< \alpha \le 2$. 
A schematic summary of the expected values for the standard long-range VM and IM (at low but finite temperature), is shown in Fig.~\ref{1suzd1and2} for both 1D and 2D.
%with those for the standard long-range IM and VM, the values of the exponent $z$ in the known models are reported in Table~\ref{1suzd1}\footnote{ In the IM the full $\alpha$-dependence of the exponent $z$ is
%\be
%z=\left \{\begin{array}{ll}
%2\,, & \quad \mbox{for } \quad \alpha>2 \\
%\alpha\,,  & \quad \mbox{for } \quad 1< \alpha \le 2.
%\end{array}
%\right .
%\label{expz}
%\ee} 
%and \ref{1suzd2} 
%in 1D\ and 2D, respectively 
%for some values of $\al$. 
%These values are chosen as to better discriminate between the VM and the IM universality classes.
%In the next section, the asymptotic exponent $z$ will be obtained from numerical simulations for values of $\alpha$ within the different classes and compared with Fig.~\ref{1suzd1}.

%\begin{table}[h!]
%\begin{tabular}{|c|c|c|}
%\hline
%\quad $\alpha$ \quad \quad & \quad VM \quad \quad & \quad IM \quad \quad \\
%\hline
% 5   & 1/2  & 1/2   \\
% 3.5 & 0 & 1/2 \\
% 2.5 & 0    & 2/3   \\
%\hline
%\end{tabular}
%\caption{Asymptotic value of the exponent $1/z$ in the VM and IM in $D=2$.}
%\label{1suzd2}
%\end{table}

%%%%%%%%%%%%%%%%%%%%%%%%%%%%%%%%%%%%%%%%%%%%%%%%%%%%
\section{Simulations of the persistent voter model} \label{simul}

Throughout this work we consider periodic boundary conditions and distances are  computed accordingly, with the values of $r$ being constrained to the interval $[1,L/2]$.
For $d=2$, in order to decrease the computational cost, the simulations were performed with a faster algorithm, where interactions only occur along the horizontal and vertical direction.
When compared with the full version, where interactions occur along all directions,
%, with the caveat that, in doing that, one has to rescale $\alpha \to \alpha+1$ to keep into account the solid angle. 
%J: this is done in both the full and fast versions, $p(r)=Ar^{-\alpha} * r$, where this last $r$ (that rescalates $\alpha$) compensates the increasing radius (as in the problem of choosing random, homogeneous points inside a circle) and using the inverse transformation method 
the two algorithms give equivalent results.
The time unit, the Monte Carlo step (MCS), involves $N$ attempts to change the state of a randomly chosen, focal agent. In each of such attempts, a single random neighbor is then chosen at a distance $r$ apart with a probability that decays as $r^{-\alpha}$. The focal agent then tries to update its opinion $S_i$ and zealot status $\theta_i$ as defined in the previous section. Notice that in the restricted application of the updating algorithm, $\theta_i$ is only updated when the chosen distance is $r=1$.

\begin{figure}[htb]
\includegraphics[width=\columnwidth]{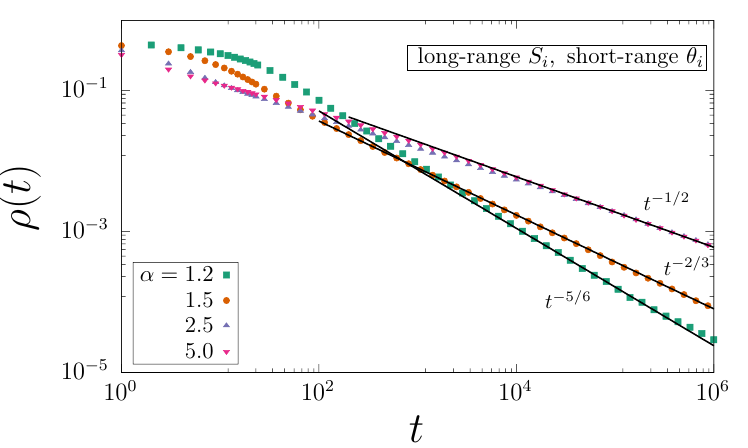}
\caption{Density of interfaces $\rho(t)$ of the PVM in $d=1$ with linear size $L=10^5$ sites for several values of $\alpha$. Averages are over $10^2$ (larger $\alpha$) or $10^3$ (smaller $\alpha$) samples.  The results are for the restricted case in which $\theta_i$ is affected by nearest-neighbors only ($r=1$). The straight lines indicate the expected exponent $1/z$ in the correspondent IM universality class (see Fig.~\ref{1suzd1and2}, top panel).}
\label{fig_LRd1-alfa}
\end{figure}

\begin{figure}[htb]
\includegraphics[width=\columnwidth]{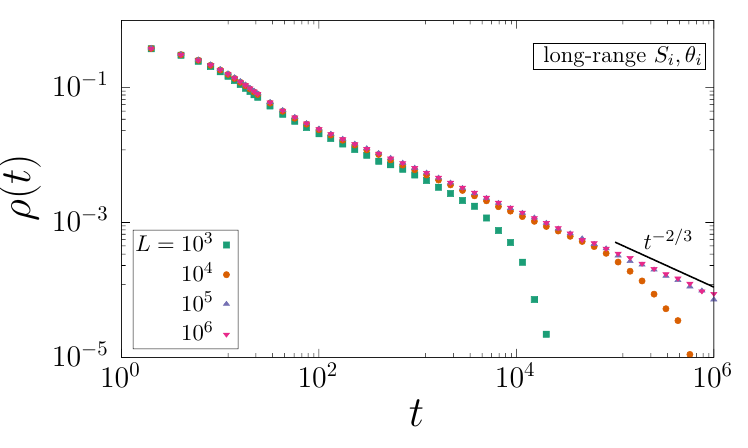}
\caption{Density of interfaces $\rho(t)$ of the long-range PVM in $d=1$ with $\alpha=3/2$  for several system sizes in the unrestricted case. The number of samples ranges from 20 (largest $L$) to 1000 (smallest $L$). There are stronger finite-size effects than in the restricted case and the Ising exponent only becomes clear at late times in the largest system.}
\label{fig_LRd1-L}
\end{figure}

Starting with the case $d=1$, we run simulations for some representative values of $\alpha$, Fig.~\ref{fig_LRd1-alfa}, and compare the PVM with the VM and the IM. %, and better discriminate between them.
Constraining $\theta_i$ to be updated only when $r=1$ makes not only the problem easier to be analytically treated but also clearly places the model within the long-range IM universality class.
Indeed, comparing the predicted exponents $1/z$ (Fig.~\ref{1suzd1and2}) with the results in Fig.~\ref{fig_LRd1-alfa}, one observes that the PVM is in  the universality class of the IM quenched to small but finite temperature (at zero temperature the exponent for the IM would be $\/z=1$ for any value of $\alpha$~\cite{Corberi_2017,PhysRevE.50.1900}).
When $S_i$ and $\theta_i$ have no restrictions whatsoever regarding the value of $r$ (Fig.~\ref{fig_LRd1-L}), there are strong finite-size effects in the WLR region $1<\alpha<2$ where the IM exponent is observed only for asymptotic times in the largest system size $L$.
The reason for that can be explained as follows. The crossover from Voter to Ising universality class is due to the action of a core of zealots inside the ordered domains, whose compact presence enhances surface tension.
In the unrestricted case such core can be partly punctured by interactions with different opinions located at large distances, outside the cluster, 
changing the zealot state of some sites and creating islands of normal voters inside the zealot core. Remember that only normal voters may change opinion, thus the restricted dynamics never presents this effective bulk noise (as defined in Ref.~\cite{OlMeSa93}). When domains start to be comparable with the system size, such islands destabilize the smaller domains promoting the breaking of the symmetry.
This is why finite-size effects are felt earlier with the unrestricted dynamics.
%In this case, the zealot core that is built up in the bulk of the domains, responsible for the emergent curvature-driven behavior, takes longer to become established and may be destabilized by interactions with different opinions located at larger distances, outside of the cluster.
A similar effect is obtained in 2D and only results for the restricted  case will be presented.

%Figure~\ref{fig.snapshots} shows, for $\alpha=2.5$ and $d=2$, snapshots of the configurations for different restrictions on the dynamics: while in the middle panel there are no restrictions on $r$ for the updating of both $S_i$ and $\theta_i$, the left (right) panels restricts the updating of $S_i$ ($\theta_i$) to  $r=1$ only. 
%Nonetheless, in all cases, the overall, qualitative picture resembles the observed during domain coarsening in the IM. 

%\begin{figure}[htb]
%\includegraphics[width=0.32\columnwidth]{PVM_conf_000200-SRS1SRZ0}
%\includegraphics[width=0.32\columnwidth]{PVM_conf_000200-SRZ0}
%\includegraphics[width=0.32\columnwidth]{PVM_conf_000200-SRZ1}
%\caption{Snapshots in $d=2$ for $\alpha=2.5$ after 200 MCS for the long-range version of the PVM. Dark colors are normal voters while light ones are zealots. The opinion and zealot variables, $S_i$ and $\theta_i$ respectively, may change upon interactions with a neighbor at a distance $r$. In the middle panel there are no restrictions on $r$. In the right panel, $\theta_i$ may change only when $r=1$ while in the left one is $S_i$ that may change only for $r=1$.}
%\label{fig.snapshots}
%\end{figure}

\begin{figure}[htb]
\includegraphics[width=\columnwidth]{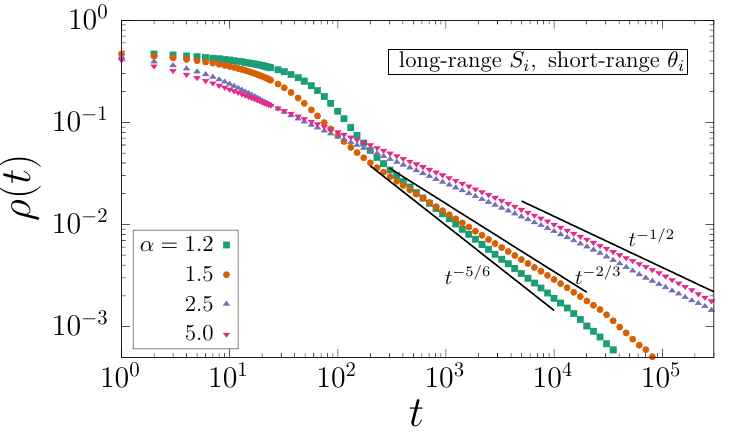}
\caption{Density of interfaces of the PVM in $d=2$, for different values of $\alpha$ and system linear size $L=10^3$, using the fast algorithm where interactions are only considered along the lattice axis. The update of $\theta_i$ is restricted to nearest-neighbors ($r=1$) only and averages are over 100-200 samples. The power-law behavior corresponding to the 2D-IM (straight lines) is well observed for all values of $\alpha$.
The down-bending observed at large times is due to finite-size effects.}
\label{fig_rho2d-J}
\end{figure}

Simulation results for $d=2$, obtained with the fast algorithm, are shown in Fig.~\ref{fig_rho2d-J}.
By comparing the measured exponent $1/z$ with the values in Fig.~\ref{1suzd1and2}, we conclude that the PVM belongs to the long-range 2D-IM universality class also in this case.
Notice that the PVM is in the universality class of the IM quenched to small but finite temperature and not to zero temperature (in the latter case the exponent for the IM would be $1/z=3/4$ for any $\alpha$~\cite{PhysRevE.103.012108,PhysRevE.103.052122}).

%\begin{figure}[htb]
%\includegraphics[width=0.9\columnwidth]{rho2d}
%\caption{Density of interfaces of the PVM in $d=2$, for three values of $\alpha$ and $L=10^3$.}
%\label{fig_rho2d}
%\end{figure}

%\begin{figure}[htb]
%\includegraphics[width=\columnwidth]{prob.touching-pvm-lr-tau15}
%\includegraphics[width=\columnwidth]{prob.touching-pvm-lr-tau25}
%\caption{Probability that the largest cluster touches two opposite sides of the lattice in both directions ($P_{11}$) or along a single direction ($P_{01}+P_{10}$). Orange (circle) symbols are for the case with $\theta_i$ dynamics constrained to nearest-neighbors while green (square) ones have no restriction whatsoever.   For $\alpha=1.5$ there seems to be a reduction in the relative number of states leading to a striped state. $L=200$.}
%\label{fig_rho2d-J}
%\end{figure}

%%%%%%%%%%%%%%%%%%%%%%%%%%%%%%%%%%%%%%%%%%%%%%%%%%%%%%%%%%
\section{Analytical results}
\label{sec.correlators}

The time evolution of the correlator $\langle\alpha_1\alpha_2 \cdots \alpha_m\rangle$, where $\alpha_i$ may be either $S_i$ or $\theta_i$, is given by
\begin{equation}
  \label{eq.evcor}
  \frac{d}{dt} \langle\alpha_1\alpha_2\cdots\alpha_m\rangle=-2\left\langle\alpha_1\alpha_2\cdots\alpha_m\sum_{i=1}^m w(\alpha_i)\right\rangle. 
\end{equation}
To simplify the notation, we define 
\begin{equation}
  \label{eq.repcor}
  C_{i,j,k,...}^{n,p,q,...}\equiv\langle S_i S_j S_k\cdots\theta_n\theta_p\theta_q\cdots\rangle,
 \end{equation}
where subscript and superscript indices are for the $S$ and $\theta$ components, respectively. 
Notice that for the restricted case considered here, the equation for the correlator with only upper (zealots) indices are the same as in the nn model, and were already analyzed in Ref.~\cite{ourpaper}.

Starting from the 1-point function $C_i$,
\begin{align}
\frac{d C_i}{d t} =  \sum_p P(\ell_p) n_p  \lf(C_i^i-C_i\ri) + \sum_p P(\ell_p) \sum_{k=1}^{n_p}  \lf(C_k-C^i_k\ri) \, .
\end{align}
Summing over $p$ and $k$ and using that $\sum _p \sum _{k=1}^{n_p} P(\ell_p)= \sum_p n_p P(\ell_p)=1$, we get
\bea
\frac{d C_i}{d t} \ = \  -C_i+C_i^i + \sum_p \, P(\ell_p) \sum_{k=1}^{n_p} \, \lf(C_k-C^i_k\ri) \, .
\eea 
Defining the discrete, long-range Laplacians
\begin{eqnarray}
\Delta _{i}&=&2 d \, \sum _p P(\ell_p) \sum_{k=1}^{n_p} C_{k}-2 d \,  C_{i} \, , \\
\Delta _{i}^i &=& 2 d \, \sum _p P(\ell_p) \sum_{k=1}^{n_p} C_{k}^i-2 d \, C_{i}^i \, ,
\end{eqnarray}
the above equation can be rewritten in the simplified form
\bea \label{eqC1}
\frac{d C_i}{d t} \ = \ \frac{1}{2 d} \lf(\Delta _{i}-\Delta _{i}^i \ri) \ .
\eea 
When the system is quenched from an uncorrelated, random initial condition, $C_i=C_j$, $C^i=C^j$ and the above equation just states that $C_i$ is constant (in that case both $\Delta _{i}$ and $\Delta _{i}^i$ are identically zero).

We thus move to the two-point correlation function for the spins, using  $w_{\rm LR}(S_j)$ as defined in Eq.~\eqref{wlrv}:
\begin{align}
\frac{d C_{ij}}{d t} &=  - \ha \left\langle S_i S_j \left[w_{\rm LR}(S_i)+w_{\rm LR}(S_j)\right]\right\rangle \nonumber \\
&+ \ha  \left\langle S_i S_j \left[\theta_i w_{\rm LR}(S_i)+ \theta_j w_{\rm LR}(S_j)\right]\right\rangle \, .
\end{align}
Apart from the $1/2$ factor, the first term also appears in the standard long-range VM (see Refs.~\cite{Corberi_20241d,corsmal2023ordering,corberi2024coarsening,corberi2024aging,Corberi_2024}). Therefore, we get
\begin{align}
\frac{d C_{ij}}{d t}  &=  - C_{i j}+ \sum _p P(\ell_p) \sum_{k=1}^{n_p} C_{j k}  \nonumber\\
&+ \ha  \left\langle S_i S_j \left(\theta_i w_{\rm LR}(S_i)+ \theta_j w_{\rm LR}(S_j)\right)\right\rangle \nonumber\\
%\end{align}
%which gives
%\begin{align}
%\frac{d C_{ij}}{d t} 
=& - C_{i j}+ \sum _p P(\ell_p) \sum_{k=1}^{n_p} C_{j k} \nonumber\\ &+  \sum _p P(\ell_p) \sum_{k=1}^{n_p} \lf(C^i_{i j}-C^i_{ j k} \ri)  \, ,
\end{align}
where we used the space-translation invariance of the system.
Grouping terms and summing over $p$ and $k$ one arrives at
\begin{align} 
\label{eqC2}
\frac{d C_{ij}}{d t}  &=  - \lf(C_{i j}-C_{ij}^i\ri )+ \sum _p P(\ell_p) \sum_{k=1}^{n_p} \lf (C_{j k}-C_{jk}^i\ri ) \nonumber\\ &=  \frac{1}{2 d} \lf(\Delta_{ij}-\Delta_{ij}^i\ri)\, ,
\end{align}
with the following definitions for the operators
\begin{align}
\Delta _{ij}&= 2d \,\sum _p P(\ell_p) \sum_{k=1}^{n_p} C_{k j}- 2d \, C_{ij} \, , \nonumber\\  \label{laplacians} \\
\Delta _{ij}^i&= 2d \, \sum _p P(\ell_p) \sum_{k=1}^{n_p} C_{k j}^i-2d \, C_{ij}^i \, . \nonumber
\end{align}

In the following we will focus on the case $d=1$, where Eqs.~(\ref{laplacians}) read
\begin{align}
\Delta _{ij}&=2 \, \sum _{\ell=1}^{N/2} P(\ell) \sum_{\sigma=\pm 1} C_{i+\sigma \ell, j}-2 \, C_{ij} \, , \nonumber \\ \label{laplaciansd1} \\
\Delta _{ij}^i&= 2 \, \sum _{\ell=1}^{N/2} P(\ell) \sum_{\sigma=\pm1} C_{i+\sigma\ell, j}^i-2 \, C_{ij}^i \,\nonumber .
\end{align}
As in the nn VM, the equations for the one- and two-point correlations basically share the same structure, i.e. the one of a diffusion equation, here made more complicated by the presence of zealots and the long-range interactions among spins. 
%However, Eq.~\eqref{eqC2} has non-trivial solutions.
Eq.~\eqref{eqC2} is not closed and despite the model not being exactly solvable, some analytical progress can be done as will be shown.
In the subsections \ref{sub.alpha3} and \ref{sub.alpha13} below, we only consider the 1D case, where
the relevant intervals are $\alpha>3$, $2<\alpha\le 3$, and $1<\alpha\le 2$, in which the model behaves differently.

\subsection{$\alpha>3$}
\label{sub.alpha3}

\begin{figure}[htb]
\includegraphics[width=\columnwidth]{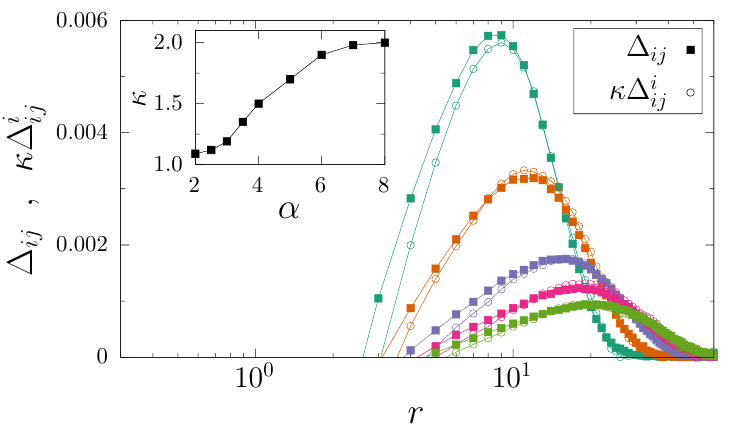}
\caption{The two Laplacians $\Delta_{ij}$ and $\kappa \Delta _{ij}^i$ are plotted against $r=|i-j|$ at different times ($t=100$, 200, 400, 600 and 800 corresponding to decreasing height of the peak) for $d=1$ and $\alpha=4$. 
Solid lines are guides to the eye. Inset: the value of $\kappa$, as a function of $\alpha$, that better obeys $\Delta_{ij}\simeq \kappa \Delta _{ij}^i$ for $t=100$.  The system linear size is $L=10^3$ and finite-size effects are still important.}
\label{fig_laplacians}
\end{figure}

For sufficiently large values of $\alpha$, the PVM with long range interactions  must behave similarly to the nn model.
For the latter, a simple geometrical argument was put forward in Ref.~\cite{ourpaper} relating the two Laplacians as $\Delta_{ij}\simeq \kappa \Delta _{ij}^i$, with $\kappa=2$, for sufficiently large $r=|i-j|$. 
We argue that a similar relation, with a possible $\alpha$-dependence in $\kappa $, still holds for sufficiently large values of $\alpha$, that we anticipate to be $\alpha >3$.
This can be checked numerically and the results are shown in Fig.~\ref{fig_laplacians}. 
Plugging this assumption into Eq.~(\ref{eqC2}), and proceeding as in Ref.~\cite{Corberi_20241d}, one arrives at a scaling form
\be
C(r,t)=f\left (\frac{r}{{\cal L}(t)}\right ),
\label{scalC}
\ee
with 
\be
{\cal L}(t)\propto t^{1/2},
\ee
and 
\be
f(x)={\rm erfc}\left (\frac{x}{x_0}\right ),
\ee
with $x_0\propto \langle \ell ^2\rangle ^{1/2}$ and $\langle \ell^2\rangle= \sum _{\ell=1}^{N/2}\ell ^2 P(\ell)$.
Notice that $\langle \ell ^2\rangle$ diverges as 
$\alpha \to 3^+$, signaling that the above solution is only valid for $\alpha>3$, as already anticipated. 
Indeed, one can check both numerically and in the simulations (see  Fig.~\ref{fig_laplacians}, inset) that, despite some finite-size effects, the results are consistent with $\kappa \to 1$ as $\alpha \to 3^+$.
%, as it is also witnessed by the fact that $\kappa (\alpha=10)\simeq 1.9$ while $\kappa (\alpha=4)\simeq 1.5$, as it can be seen in. 

\subsection{$1<\alpha \le 3$}
\label{sub.alpha13}

\begin{figure}[htb]
\includegraphics[width=\columnwidth]{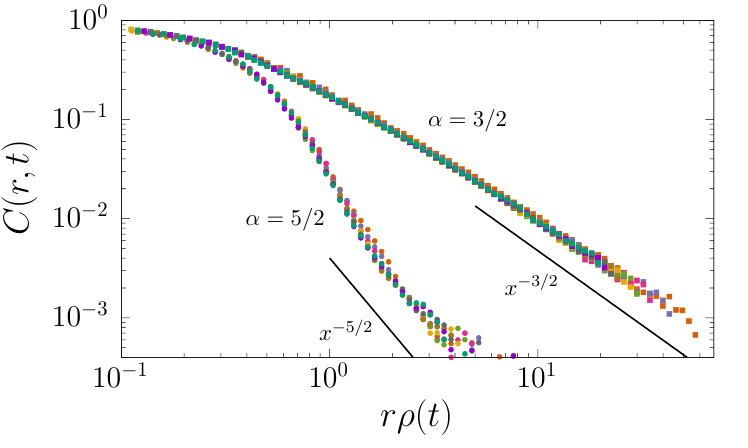}
\caption{The correlation $C(r,t)\equiv C_{ij}$ is plotted against $r\rho (t)$ for $d=1$, $\alpha=1.5$ and 2.5 at times $t=20,\ldots, 100$ (different colors). The system linear size is $L=1000$ and averages are taken over $10^4$ samples.}
\label{scal_corr}
\end{figure}

Within the range $1<\alpha \le 3$, the VM displays scaling, Eq.~(\ref{scalC}), with
\be
f(x)\propto x^{-\alpha},
\label{fx}
\ee
for sufficiently large $x\gtrsim 1$.
The numerical results in  Fig.~\ref{scal_corr} suggests that the same applies in the PVM, what will later be checked for consistency.
%and is shown to be correct by the numerical results contained in Fig.~\ref{scal_corr}.
With this scaling, Eq.~(\ref{eqC2}) reads
\be
\frac {\partial C_{ij}}{\partial t}=\frac{1}{2}(\Delta _{ij}-\Delta_{ij}^i)=-xf'(x)\frac{\dot {\cal L}}{{\cal L}}\propto \frac{x^{-\alpha}}{{\cal L}^z}\,,
\label{dCscal}
\ee
where $\dot {\cal L}\equiv d{\cal L}(t)/dt$ and we have assumed an algebraic growth ${\cal L}(t)\propto \rho^{-1}\propto t^{1/z}$.  
%Furthermore, using Eqs.~(\ref{scalC}) and (\ref{fx}) one has
%\be
%\Delta _{ij}\propto L^{-2}x^{-\alpha -2}.
%\ee

As we have seen in the previous section, $\kappa\to 1$ when  $\alpha \to 3^+$, signaling that the two Laplacians in Eq.~(\ref{dCscal}) differ only by subleading terms, which are difficult to calculate. 
Because of that, the exponent $z$ cannot be easily computed explicitly. 
However, in the following,
we only provide a physically-oriented, heuristic argument to determine it.
The typical size $\xi$ over which correlations spread can be defined as
\be
\xi (t)=\frac{\sum_{r=1}^{L/2} r\,C(r,t)}{\sum_{r=1}^{L/2} C(r,t)}.
\ee
Letting $L\to \infty$, this quantity is finite only if $\alpha >2$. We can expect that, in this case, when ${\cal L}$ has grown much larger than $\xi$, namely for large
times, the growth mechanism remains the same as in the short range case, at least in its universal features. This would imply $z=2$, for any $\alpha >2$. It remains to describe what occurs for $1<\alpha \le 2$.
Notice that in the limit $\alpha \to 1^+$, %the {\it interaction} 
$P(r)$ becomes non-integrable. This means that single interfaces are aware of the whole system configuration and, hence, move deterministically in the most favorable direction, implying a ballistic behavior with $z=1$. This is observed both in the IM~\cite{Corberi_2017,PhysRevE.50.1900} and in the VM~\cite{Corberi_20241d} in $d=1$. Assuming that the growth exponents are continuous functions of $\alpha$ (which is always observed in coarsening models), the simplest ansatz interpolating between $z(\alpha=2)=2$ and $z(\alpha=1)=1$ is $z=\alpha$. The above argument reproduces the IM-like behavior %~(\ref{expz}) 
observed in numerical simulations of the PVM (see Sec.~\ref{simul}).

%%%%%%%%%%%%%%%%%%%%%%%%%%%%%%%%%%%%%%%%%%%%%%%%%%%%%%%%%%%%%%
\section{Conclusions} 
\label{conc}

We have studied the coarsening kinetics of a long-range variant of the PVM, in which agents can be either normal voters or transient zealots and whose interactions decay with a power-law of the distance, $r^{-\alpha}$. 
This version of the PVM is interesting not only because it is more approachable analytically, but mainly because it has the same overall behavior of the original version. Allowing the confidence level to change only via nn interactions reduces the finite-size effects in the simulations and makes the analytical approach possible. By means of numerical simulations in $d=1$ and 2 for various values of $\alpha$, we have shown that the long-range PVM belongs to the same universality class as the long-range IM (with analogous interactions), as already observed in the nn case \cite{ourpaper}. Indeed, the interface density decays with growth exponents consistent with those of the long-range IM quenched to a small but finite temperature, $T \neq 0$. Furthermore, for the one-dimensional case we have provided an analytical treatment based on one- and two-point correlation functions, reproducing the $\alpha$-dependence of the correlation length and of the correlation function. These results seem to confirm that the introduction of opinion inertia qualitatively modifies the dynamics of the VM, reducing interfacial noise and inducing an IM curvature-driven growth mechanism even in the presence of long-range interactions.

Finally, it is worth emphasizing a point. In Refs.~\cite{Kleinberg2000,SeBaBi02,ChSe06}, the authors considered networks embedded in Euclidean space, where long-range links between pairs of sites are added with a probability that decays as a power law of their Euclidean distance, $P(r)\sim r^{-\alpha}$. One might therefore be tempted to compare that setting with the one considered in the present paper. However, the two approaches are fundamentally different. In Refs.~\cite{Kleinberg2000,SeBaBi02,ChSe06}, the power-law probability is used to build the topology of the embedded network itself. By contrast, in our model, while the topology is fixed, coinciding with that of a regular lattice, the interaction network is not, fluctuating along the time. Indeed, the power-law distribution enters only through the dynamical update rule that attempts to change the opinion or the zealot state of a focal agent 
%copies the state of 
by another agent, chosen at a distance $r$ using the power-law distribution.
%, provided that the selected agent is not a zealot.
This distinction has important physical consequences, since the two systems do not belong to the same universality class. This can be inferred, for instance, by comparing the behavior found in our work with the results reported in Ref.~\cite{BiSe11}, where the corresponding case is referred to as Random Model B.

\begin{acknowledgments}
Partial funding by the Brazilian Conselho Nacional de Desenvolvimento Científico e Tecnológico (CNPq), Grants 316628/2021-2 (JJA), 443517/2023-1 (JJA), 313621/2026-8 (JJA) and 402487/2023-0 (WGD and JJA), is acknowledged.

\end{acknowledgments}
%%%%%%%%%%%%%%%%%%%%%%%%%%%%%%%%%%%%%%%%%%%%%%%%%%%%%%%%%%%%%%
%\bibliographystyle{apsrev4-2}   
%\bibliography{zealots} 

\end{document}